\documentclass[fleqn,usenatbib]{mnras}

\usepackage{newtxtext,newtxmath}

\usepackage[T1]{fontenc}
\usepackage{ae,aecompl}

\usepackage{graphicx}	
\usepackage{amsmath}	

\title[Nuclear burning in an AMS disk]{Nuclear burning in an accretion flow around a stellar-mass black hole embedded within an AGN disk}
\author[Z. Tang et al.]{
Zifan~Tang,$^{1}$
Yang~Luo,$^{1}$\thanks{E-mail: luoyang@ynu.edu.cn (YL)}
Jian-Min~Wang,$^{2,3,4}$
\\
$^{1}$Department of Astronomy, Yunnan University, Kunming, Yunnan 650091, China\\
$^{2}$Key Laboratory for Particle Astrophysics, Institute of High Energy Physics,
Chinese Academy of Sciences, 19B Yuquan Road, Beijing 100049, China\\
$^{3}$School of Astronomy and Space Sciences, University of Chinese Academy of Sciences, 
19A Yuquan road, Beijing 100049, China\\
$^{4}$National Astronomical Observatory of China, 20A Datun Road, Beijing 100020, China
}

\date{Accepted XXX. Received YYY; in original form ZZZ}

\pubyear{2024}

\begin{document}
\maketitle
\label{firstpage}

\begin{abstract}

A stellar-mass black hole, embedded within the accretion disk of an active galactic nuclei (AGN), has the potential to accrete gas at a rate that can reach approximately $\sim 10^9$ times the Eddington limit. This study explores the potential for nuclear burning in the rapidly accreting flow towards this black hole and studies how nucleosynthesis affects metal production. Using numerical methods, we have obtained the disk structure while considering nuclear burning and assessed the stability of the disk. In contrast to gas accretion onto the surface of a neutron star or white dwarf, the disk remains stable against the thermal and secular instabilities because advection cooling offsets the nuclear heating effects. The absence of a solid surface for a black hole prevents excessive mass accumulation in the inner disk region. Notably, nuclear fusion predominantly takes place in the inner disk region, resulting in substantial burning of $\rm ^{12}C$ and $\rm ^{3}He$, particularly for black holes around $M = 10\, M_\odot$ with accretion rates exceeding approximately $\sim 10^7$ times the Eddington rate. The ejection of carbon-depleted gas through outflows can lead to an increase in the mass ratio of oxygen or nitrogen to carbon, which may be reflected in observed line ratios such as $\rm N\, V/C\, IV$ and $\rm O\, IV/C\, IV$.  Consequently, these elevated spectral line ratios could be interpreted as indications of super-solar metallicity in the broad line region.

\end{abstract}




\section{Introduction}

An accretion disk surrounding a compact object can release gravitational energy and is believed to be responsible for various cosmic phenomena, such as active galactic nuclei (AGN) or X-ray sources \citep{ho08,abramowicz13,yuan14,netzer15,padovani17}. Many researchers have developed theoretical models of these accretion disks that operate at different accretion rates \citep{shakura73,abramowicz88,narayan94}. The radiative cooling emitted from the surface of the accretion disk can reduce the disk thickness until viscous heating balances the radiative loss. When this balance is not achieved, especially in the outer regions of the disk, the self-gravity of the disk becomes significant, potentially leading to local instability \citep{paczynski78, abramowicz84,lodato07,wang11}. 

As instability develops, it can result in fragmentation through gravitational instability \citep{shlosman87,goodman03, goodman04,lodato07,collin08,wang11,mapelli12}. The fragments may collapse further, eventually forming stars \citep{shlosman89b, goodman04, cantiello21}. 

Alternatively, stars within nuclear star clusters may be captured by the accretion disk \citep{artymowicz93,wang24}. Observations of stellar clusters around the galactic center and the notably top-heavy initial mass function \citep[e.g.,][]{nayakshin05,paumard06,bartko10,zhu18,nayakshin18,neumayer20,schodel20,jia23} suggest star formation may have occurred in the Galactic center several million years ago. Subsequently, stars and compact objects within these clusters might have slowly drifted towards the SMBHs due to dynamical friction with surrounding gas and stars \citep{gerhard01}. The rate of this drift is affected by the distribution of orbit eccentricity of these stars \citep{collin08, kennedy16}. Furthermore, the possible connection between the spatial locations of quasar SDSS J1249+3449 and the gravitational wave event GW190521 \citep{graham20} hints at the possibility of forming massive black hole binaries in AGN disks, where the dense environments promote binary mergers \citep{samsing22}. It appears likely that stars or compact objects could reside within the AGN disk.

Whether formed due to disk instability or captured from the cluster, these stars or compact objects can continue growing through gas accretion \citep{dittmann20,wang21,ali-dib23,wang24}.



Unlike field stars in their host galaxies, stars located in self-gravitating AGN disks can accrete gas at rates significantly exceeding the Eddington limit, primarily because of the dense environment of the AGN disk \citep{wang21}. An estimation of the accretion rate towards the central object suggests that the Eddington ratio could reach values around $\sim 10^9$ \citep{wang21}. For simplicity, we refer to these rapidly accreting stars as accretion-modified stars (AMSs) \citep{wang21, wang21b}. It is important to note that the central star could be a normal star or a compact object. Certainly, by examining a star age and the AGN lifetime \citep{schawinski15, king15}, it is still plausible that the compact object could originated from surrounding star clusters and was subsequently captured by the AGN disk.

For an AMS, assuming a central black hole (BH), with an accretion rate as high as $\sim 10^9 \dot{M}_{\rm edd}$, where $\dot{M}_{\rm edd}$ is the BH Eddington rate, is there a potential for initiating nuclear burning within the accretion disk of this AMS?

The sensitivity of nuclear reaction rates to temperature introduces the concept of ignition or threshold temperature \citep{prialnik09}. When this threshold is surpassed, the reaction rates increase significantly. For hydrogen burning, the threshold temperature for the proton-proton chain is around $4\times 10^6$ K, and for the CNO bi-cycle process, it is approximately $1.5\times 10^7$ K. In the numerical simulations \citep{kitaki18}, it is implied that for supercritical accretion disks, the temperature within the disk mid-plane can be approximated by a formula $T = 3.85\times 10^7 {\rm K} \ (M/{M_{\odot}})^{-0.24}\, (\dot{M}/\dot{M}_{\rm edd})^{0.24}\,(r/2r_{\rm g})^{-0.54} \label{eq-dep-tgas}$, where $M$, $\dot{M}$, and $c$ are the BH mass, accretion rate, and the speed of light, respectively. The gravitational radius $r_{\rm g}$ is defined as $r_{\rm g} \equiv GM/c^2$. Consequently, the temperature in the accretion disk, by assuming supercritical accretion, could potentially reach the threshold for nuclear burning. 

The timescale of a nuclear reaction is inversely proportional to the reaction rate \citep{iben12}. For an estimate of this timescale, we consider a density of $\rho \sim 1\ \rm g\,cm^{-3}$ and a temperature of $T \sim 1.6\times10^7\ \rm K$. Refer to Figure \ref{fig:disk_profile}, the densities at various accretion rates have already increased to approximately an order of magnitude greater than $\sim 1\ \rm g\,cm^{-3}$ near the inner disk area.  Using these parameters, we determine that the lifetime of the isotope $\rm ^{16}O$, which has the longest lifetime among the nuclei in the CNO bi-cycle, to be approximately $\sim 10^{-4}\ \rm s$ \citep[see table 6.10.2 in][]{iben12}. This timescale is already shorter than the disk viscous timescale. As the disk accretion rate increases, the density and temperature near the disk center could also increase, and thus the timescale, which also depends on the disk radius, might become even shorter.

Therefore, based on the above estimation of the disk temperature and the nuclear timescale, it is expected that the accreting material in the AMS disk may experience phenomena like hydrogen burning or alpha captures.

Observational phenomena, such as novae \citep[e.g.,][]{shen07,hachisu19} or type I X-ray bursts \citep[e.g.,][]{narayan03,zamfir14,galloway17,galloway21}, are thought to be linked to nuclear reactions that occur in an unstable fashion. It is hypothesized that hydrogen shell burning or thermonuclear explosions may occur on the surface of white dwarfs or neutron stars \citep[for a review, see][]{lewin93, galloway17}. Are thermonuclear explosions possible in a super-critical accretion disk? Can nuclear burning affect disk stability or cause variations in the accretion rate? Or could we expect any type of observational flare, caused by these instabilities, in the light curves ?

A pertinent question is whether nucleosynthesis can significantly modify the abundance of heavy elements in the disk. Observations suggest that the metallicity of the broad line region (BLR) in high-redshift quasars exceeds solar levels, inferred from specific emission line ratios \citep{nagao06,shin17,lai22,huang23}. This super-solar metallicity might result from star formation in the nuclear region of the AGN host galaxy, where these stars subsequently enhance the metal content of quasars (see the review paper of \citet{hamann99}). Furthermore, if the energy generated by nuclear reactions within the accretion disk of AMS is considerable, we might also expect some metal contributions from the outflow within the accretion disk. 

Research on nuclear reactions and nucleosynthesis in accretion flows around black holes has been previously examined by various scholars \citep{taam85,chakrabarti87,arai92,mukhopadhyay00,mukhopadhyay01,hu08}. However, studies on accretion rates as high as $\sim 10^9 \dot{M}_{\rm edd}$ have not yet been performed. Furthermore, earlier studies had to assume very inefficient angular momentum transport, with $\alpha$ viscosity levels as low as $\sim 10^{-10} - 10^{-6}$ \citep{arai92}, or ignore the effect of advection cooling, which could be important for the supercritial accretion disk \citep{taam85}.

In this paper, we revisit the examination of nucleosynthesis within the fast accretion flow towards an AMS, with a focus on the possibility of nuclear burning. The paper is organized as follows. Section 2 provides a brief overview of the equations used to model the disk structure surrounding a black hole. Section 3 delves into the thermodynamic conditions within the disk, presents the findings on the disk structure incorporating nuclear burning, and assesses the stability of the disk. Section 4 explores nucleosynthesis in the disk and considers the potential implications for metal enrichment within the AGN disk. The paper concludes with final remarks in the last section.


\section{Description of the model}

At very high mass accretion rates, particularly those at or exceeding the Eddington limit, the optical depth becomes large, causing radiation to be trapped within the inflowing gas \citep{abramowicz88}. This supercritical disk is represented by the slim disk model, which is a one-dimensional accretion disk that incorporates the photon-trapping effect as the advection of photon entropy in the energy equation \citep{abramowicz88, kato08}. In the following, we adopt the formalism of the super-critical slim disk structure and include the impact of nuclear reactions. 

\subsection{Structure of the disk}

In this study, we focus on a steady and axisymmetric accretion flow, disregarding time-dependent properties. The disk model is established through the solution of the energy and momentum balance equations with specified values of the accretion rate $\dot{M}$, the black hole mass $M$, and the viscosity coeffcient $\alpha$.

Cylindrical coordinates ($r, \varphi, z$) are employed, centered around the black hole. The hydrodynamic equations, involving mass conservation, radial momentum equation, angular momentum conservation, and vertical hydrostatic balance along with the energy equation, can be expressed as follows.

\begin{eqnarray}
\label{mass}
    \dot{M}
     &=& -2\pi r v_{r} \Sigma, 
\\
\label{r-mom}
     v_r \frac{dv_r}{dr}+ \frac{1}{\Sigma}\frac{d\Pi}{dr}
     &=& r (\Omega^2-\Omega_{\rm K}^2) 
         - \frac{\Pi}{\Sigma}\frac{d \ln \Omega_K}{dr},  
\\
\label{ang-mom}
    \dot{M} (\ell-\ell_{\rm in}) 
     &=& -2 \pi r^2 T_{r \varphi}, 
\\
\label{hydro}
   (2N+3)
   \frac{\Pi}{\Sigma} 
     &=& H^2 \Omega_{\rm K}^2, 
\\
\label{energy}
   Q_{\rm vis} + Q_{\rm nuc}
     &=& Q_{\rm adv} + Q_{\rm rad}.
\end{eqnarray}

Here, $\Sigma$, $v_{\rm r}$, $\Omega$, $\Omega_{\rm k}$, $\ell$, $\ell_{\rm in}$, and $H$ represent surface density, radial velocity, angular velocity, Keplerian angular velocity, specific angular momentum, specific angular momentum at the inner edge of the disk, and disk scale height, respectively.  

The $r$-$\varphi$ component of the height-integrated viscous stress tensor is denoted by $T_{r\varphi}$, related to the total pressure by $T_{r\varphi} = -\alpha \Pi$, where $\Pi$ stands for the height-integrated total pressure $p$ defined as $\Pi \equiv \int p dz$, and $\alpha$ represents the viscosity parameter. The model is vertically integrated, and the surface density is determined as $\Sigma \equiv \int \rho dz$, where $\rho$ is the density at the disk mid-plane. A polytropic relationship $p \sim \rho^{1+1/N}$ is assumed in the vertical direction, with $N = 3$ as the polytropic index.

The total heating rate comes mainly from two sources: viscous heating $Q_{\rm vis}$ and nuclear burning $Q_{\rm nuc}$. The viscous heating rate is determined by the shear $T_{r \varphi}$,  and can be written as:
\begin{equation}
\label{qvis}
Q_{\rm vis} = - r T_{r \varphi} \frac{d\Omega}{dr}.
\end{equation}

The cooling in the disk is mostly due to advection and radiation. The advective cooling is determined approximately as

\begin{equation}
\label{qadv}
Q_{\rm adv} \approx \frac{\dot{M}}{2 \pi r^2} \frac{\Pi}{\Sigma} \xi.
\end{equation}

The variable, $\xi$, which is of order unity, is a dimensionless quantity given by

\begin{equation}
\xi = -(A_{\xi}+0.5)\frac{d\ln{\Pi}}{d\ln{r}} + (A_{\xi}+1.5)\frac{d\ln{\Sigma}}{d\ln{r}} + \frac{d\ln{\Omega_{\rm k}}}{d\ln{r}},
\end{equation}

where $A_{\xi} = 3(1-\beta) + \beta/(\gamma -1)$ and $\beta$ is the ratio of the gas pressure to the total pressure. $\gamma$ is the ratio of the generalized specific heats. 

The disk is assumed to be optically thick, and the radiative cooling, related to the temperature $T$ at the disk mid-plane, is equal to
\begin{equation}
\label{qrad}
Q_{\rm rad} = \frac{8 a_{\rm r} c T^4}{3 \kappa \rho H},
\end{equation}
where $a_{\rm r}$ is the radiation density constant. We adopt the Rosseland-mean opacity $\kappa=0.4+0.64\times 10^{23}\rho T^{-3}$ ~[cm$^{2}$g$^{-1}$].

\subsection{Nuclear reactions}

Nuclear reactions occurring in the mid-plane of the disk have the potential to generate significant energy that can influence the dynamics of the disk. Typically, for a low mass accretion rate, the rate of nuclear reactions is considered to be insignificant as a result of the low density and temperature of the material. However, in cases of rapid accretion, where the material is denser and hotter, nuclear reactions play a crucial role \citep{taam85, arai92}.

In order to simplify steady-state disk modeling due to computational limitations, two types of hydrogen burning reactions are taken into account: the proton-proton chain reaction and the CNO bi-cycle reaction. Furthermore, helium burning is considered through the three-alpha reaction.

 The specific energy-generation rates per unit mass in unit of $\rm erg\,g^{-1}\,s^{-1}$ of these reactions, adopted from \citet{iben12} (see their equations 6.9.38, 6.10.9, and 16.2.50), are
\begin{eqnarray}
    \epsilon_{\rm pp} 
    &\approx & X_1^2 \rho/T_6^{2/3} \rm exp(14.74-33.81/T_6^{1/3}) \beta_{\rm n}, \label{equ:nuc1}
\\
    \epsilon_{\rm CNO} 
    &\approx & X_{N}X_1 \rho/T_6^{2/3} \rm exp(64.232-152.299/T_6^{1/3}), \label{equ:nuc2}
\\
    \epsilon_{3 \alpha} 
    &\approx & X_4^3 \rho^2/T_8^3 \rm exp(31.12-44/T_8),\label{equ:nuc3}
\end{eqnarray}

where $\beta_{\rm n} = 1+J_{\rm k}(0.9576-0.4575\Gamma)/(1+\Gamma)$. Here, $T_6$ and $T_8$ represent the temperature in units of $10^6$ and $10^8$ K, respectively. The parameter $\beta_{\rm n}$ is roughly on the order of unity, whereas $J_{\rm k}$ varies with temperature. At low temperatures, $J_{\rm k}$ is quite small; however, when the temperature surpasses $10^{7-8}$\, K, $J_{\rm k}$ approaches a value of close to one. More details on the calculations of $J_{\rm k}$ and $\Gamma$ could be found in Chapter 6.9, of \citet{iben12}. In the calculations, we assume the mass fraction of hydrogen $X_1 = 0.71$, the mass fraction of helium $X_4 = 0.21$ and $X_{N} = 0.01$. The energy generation rats are very sensitive to the temperature, as indicated from the exponential part of these equations.

Nuclear processes such as electron-positron pair annihilation, bremsstrahlung, plasmon decay, and URCA process are not included in this study. By examining the density and temperature profiles presented in Section \ref{sec:results}, we determined that the energy contributions from these mechanisms remain negligible. These processes might become significant for accretion disks where neutrino cooling predominates, but our accretion rates are still below those found in neutrino-cooled disks \citep[e.g.,][]{popham99, janiuk07,luo13}. Moreover, spallation reactions involving high-energy particles with projectile energies exceeding MeV or GeV are also not considered in this work.

After we obtain the disk structure, we utilize the disk density and temperature in a post-process way to calculate the nucleosynthesis. In this analysis phase, we numerically evaluate the nuclear reaction rates and the nuclear heating term through the publicly accessible code, using a widely accessible alpha chain reaction network \footnote{https://cococubed.com/code\_pages/burn\_helium.shtml}. This reaction network, as described by \citet{weaver78} and \citet{timmes99}, incorporates 21 isotopes within an alpha chain spanning from helium to nickel, including isotopes that facilitate various forms of hydrogen burning such as proton-proton chains and steady-state CNO bi-cycles. 
This network is also the workhorse network used by the MESA code, which is used to solve the stellar evolution equations \citep{paxton11}.

\subsection{Equation of state}

We assume the equation of state adopted from \citet{paczynski83}, where the photons are assumed to follow the blackbody formula. The pressure is calculated as
\begin{equation}
    p = \frac{1}{3} a_{\rm r} T^4  + \frac{\rho k_{\rm b} T }{\mu \mathrm{m_p}}+ K\rho^{5/3},
    \label{21}
\end{equation}
where $k_{\rm b}$, $m_\mathrm{p}$, $\mu$ and $K$ are the Boltzmann constant, mass of the proton, total mean molecular weight, and $K = 3.12\times10^{12}(1+X_1)^{5/3}$.

\subsection{Black hole rotation}
\label{sec:kerr}

The structure of the relativistic accretion disk is calculated using the correction factors derived in \citet{riffert95} for a Kerr black hole.  The BH spin is parameterized by a dimensionless specific angular momentum, $a$, and the following functions of $a$, $M$, and radial coordinate $r$ are introduced:

\begin{eqnarray}
    A 
    &=& 1 -{2 G M \over c^{2} r} + \left({GM a \over c^{2}r}\right)^{2},\\
    B 
    &=& 1 - {3 G M \over c^{2} r } + {2 a \left({GM \over c^{2} r}\right)^{3/2}},\\
    C 
    &=& 1 - 4 a \left({GM \over c^{2} r}\right)^{3/2} + 3 \left({GM a \over c^{2}r}\right)^{2},\\
    D &=& {1 \over 2 \sqrt{r} } \int^{r}_{r_{\rm ms}} {{x^{2}c^{4} \over G^{2}} - 6{M x c^{2} \over G} + 8a\sqrt{M^{3}x c^{2} \over G} - 3 a^{2}M^{2} \over \sqrt{x}\left({x^{2}c^{4} \over G^{2}} - 3 {M x c^{2} \over G} +2a\sqrt{M^{3}x c^{2} \over G}\right)} dx.
\end{eqnarray}

The above coefficients approach unity at large radii and for small spin parameter. The inner boundary of the disk is located at the last stable circular orbit $r_{\rm ms}$, depending  on the black hole spin \citep{bardeen72}:
\begin{equation}
r_{\rm ms} = {GM \over c^{2}}(3+z_{2} + \sqrt{(3-z_{1})(3+z_{1}+2 z_{2})}),
\label{eq:rms}
\end{equation}
where $z_{1}=1+(1-a^{2})^{1/3}((1+a)^{1/3}+(1-a)^{1/3})$, 
$z_{2} = (3 a^{2}+z_{1}^{2})^{1/2}$.


The increase of gravity in the close vicinity of the rotating black hole in 
consequence leads to a smaller disk height, which will be given by the hydrostatic 
equilibrium:
\begin{equation}
H = { \sqrt{(2N+3)} c_{\rm s} \over \Omega_{\rm k}} \sqrt {B \over C}.
\end{equation}
where the sound speed $c_{\rm s}$ is defined as $c_{\rm s}^2 \equiv \Pi/\Sigma$.

The angular velocity in the disk around a spinning black hole is given by:
\begin{equation}
\Omega = {c \over r_{\rm g} \left(\left({r/r_{\rm g}}\right)^{3/2}+a\right)}
\label{eq:omegad}
\end{equation}

The viscous shear will be modified as:
\begin{equation}
T_{r\varphi}=-\alpha \Pi {A \over \sqrt{BC}}.
\end{equation}

Consequently, the viscous heating term in the energy balance equation will be
affected by the above correction factors and modified as $Q_{\rm vis} D/B$. This factor $D/B$ is equal to zero at the radius $r_{\rm ms}$ and approaches unity at large radii.  

\begin{figure*}
    \center 
    \includegraphics[width=0.98\textwidth]{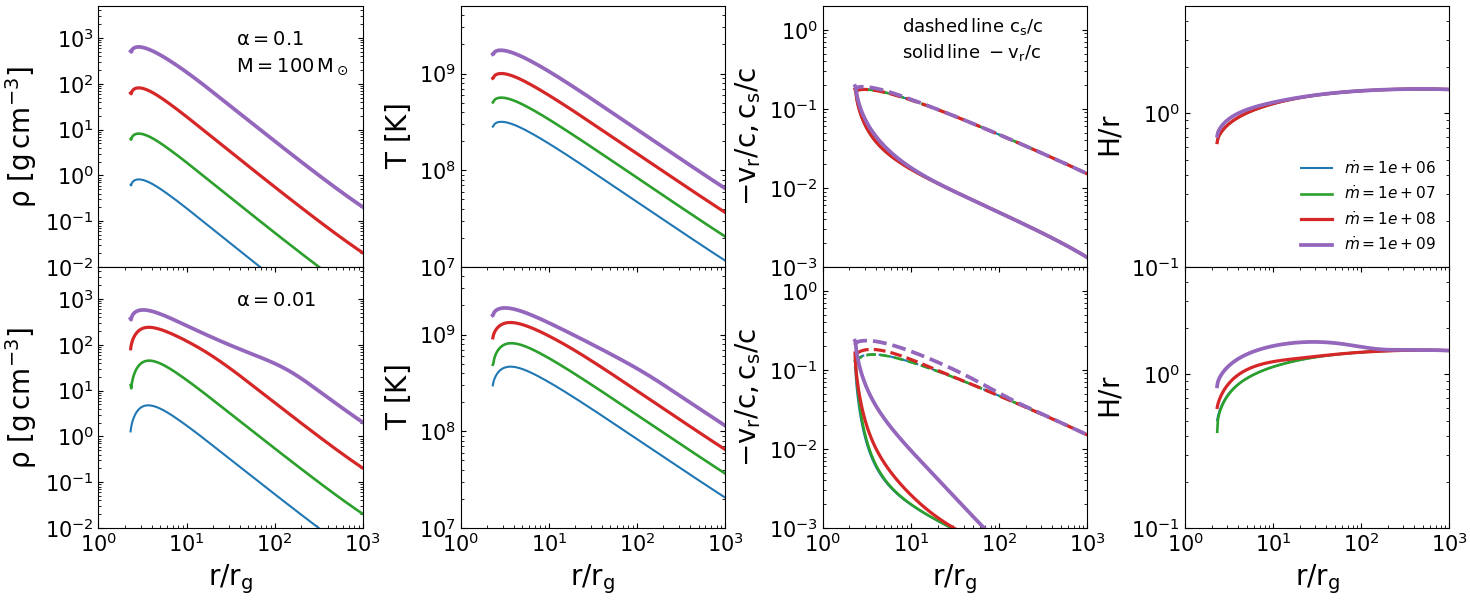}
    \caption{ The disk structure as function of radius, for models with a black hole mass of $M = 100\, M_\odot$. The panels from left to right display profiles for gas density $\rho$, temperature $T$, radial/sound velocity ($-v_{\rm r}/c, c_{\rm s}/c$) in units of $c$, and disk height aspect ratio $H/r$. Different colored lines represent models with varying accretion rates. The upper panels illustrate models with $\alpha = 0.1$, while the lower panels corresponds to $\alpha = 0.01$.}
    
    \label{fig:disk_profile}
\end{figure*}

\begin{figure}
    \center 
    \includegraphics[width=0.46\textwidth]{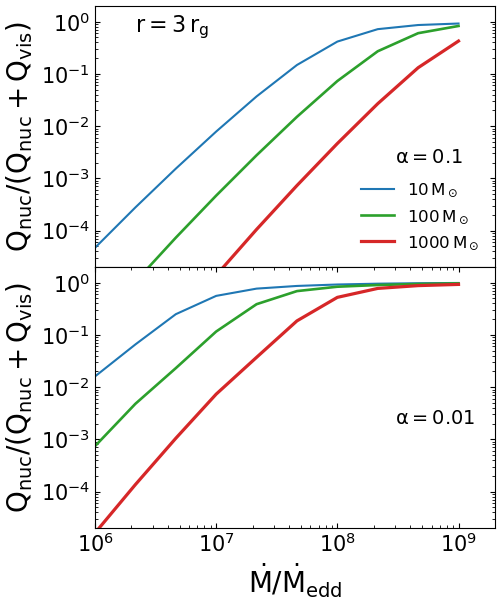}
    \caption{The contribution of nuclear heating to the total heating rate, as function of accretion rates. Ratios are adopted at disk radius of $r = 3 r_{\rm g}$. The colored lines represent models with different black hole mass.}

    \label{fig:ratio_qnuc}
\end{figure}

\begin{figure}
    \center 
    \includegraphics[width=0.46\textwidth]{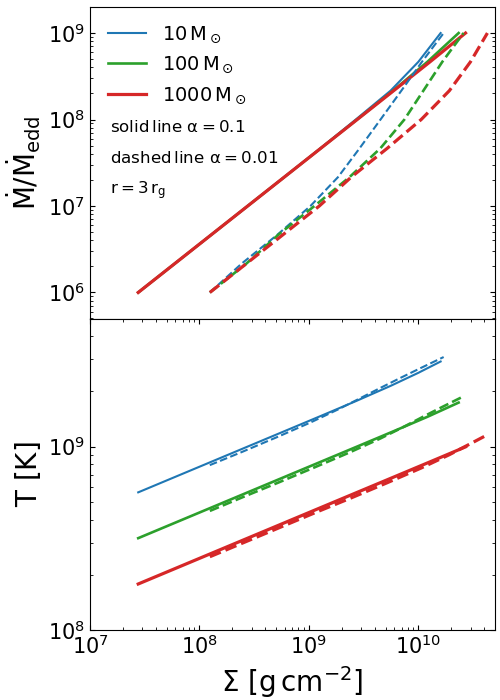}
    \caption{The $\dot{M} - \Sigma$ plane (upper panel) and the $T - \Sigma$ plane (bottom panel) at $r = 3 r_{\rm g}$ for various black hole masses. Models with $\alpha = 0.1$ and $\alpha = 0.01$ are depicted as solid and dashed lines, respectively.}

    \label{fig:instablility}
\end{figure}

\subsection{Numerical method}
\label{sec:numeric}

We consider a super-critical accretion flow onto a black hole by introducing mass from the outer boundary at a constant rate of $\dot{M}$. The inner region might involve significant nuclear burning, where the self-similar method may not accurately capture the profiles. Instead of employing the self-similar approach to solve the disk structure, we opt for a numerical method. 

Through a series of algebraic manipulations of equations \ref{mass} - \ref{energy}, we obtain the following set of two ordinary differential equations for $v_{\rm r}(r)$ and $\Pi(r)$


\begin{eqnarray}
\label{two_ode}
    &[(A_{\xi}+0.5)\frac{v_{\rm r}^2}{c_{\rm s}^2}-(A_{\xi}+1.5)]\Gamma_{v_{\rm r}} = q_{\rm vis} + q_{\rm nuc} - q_{\rm rad} + \nonumber \\
    &(A_{\xi}+1.5) - \Gamma_{\rm k} + (A_{\xi}+0.5)(\frac{r^2\Omega^2-r^2\Omega_{\rm k}^2}{c_{\rm s}^2} - \Gamma_{\rm k}), \\
    & \Gamma_{\rm \Pi}  = \frac{r^2\Omega^2-r^2\Omega_{\rm k}^2}{c_{\rm s}^2} - \Gamma_{\rm k} - \frac{v_{\rm r}^2}{c_{\rm s}^2}\Gamma_{v_{\rm r}}.
\end{eqnarray}

In this context, the parameters $\Gamma_{v_{\rm r}}, \Gamma_{\rm k}, \Gamma_{\rm \Pi}$ are associated with the derivatives $\rm dlnv_{\rm r}/dlnr$, $\rm dln\Omega_{\rm k}/dlnr$, and $\rm dln\Pi/dlnr$, accordingly. Meanwhile, the parameters $q_{\rm vis}, q_{\rm nuc}, q_{\rm rad}$ determine the relative magnitudes of Q$_{\rm vis}$, Q$_{\rm nuc}$, and Q$_{\rm rad}$ in comparison to $\dot{M} c_{\rm s}^2/2 \pi r^2$. We solve the pair of equations to determine the values of $\Gamma_{v_{\rm r}}$ and $\Gamma_{\rm \Pi}$, after which we update $v_{\rm r}(r)$ and $\Pi(r)$ accordingly. 

The flow structure is calculated from the outer edge located at $r = 10^4 r_{\rm g}$ to $r_{\rm ms}$ passing through the transonic point. At the outer boundary, we enforce the disk angular momentum to be Keplerian, with the disk height aspect ratio $H/r \sim 0.1$. At the inner boundary $r_{\rm ms}$, we consider torque-free conditions. We adjust $l_{\rm in}$, the specific angular momentum at $r_{\rm ms}$, to satisfy the regularity condition at the transonic radius, ensuring the accretion flow undergoes a sonic transition in the inner region. The location of the sonic radius could be determined by the outer edge and also the choice of $\alpha$ \citep[for a discusion of the sonic radius, see][]{lu99}. 

Subsequently, we utilize the obtained density and temperatures to calculate nucleosynthesis in the code. During the nucleosynthesis process, we assume that the element abundance in the accretion flow is similar to that of the solar atmosphere \citep{lodders03}.

\begin{figure*}
    \center 
    \includegraphics[width=0.9\textwidth]{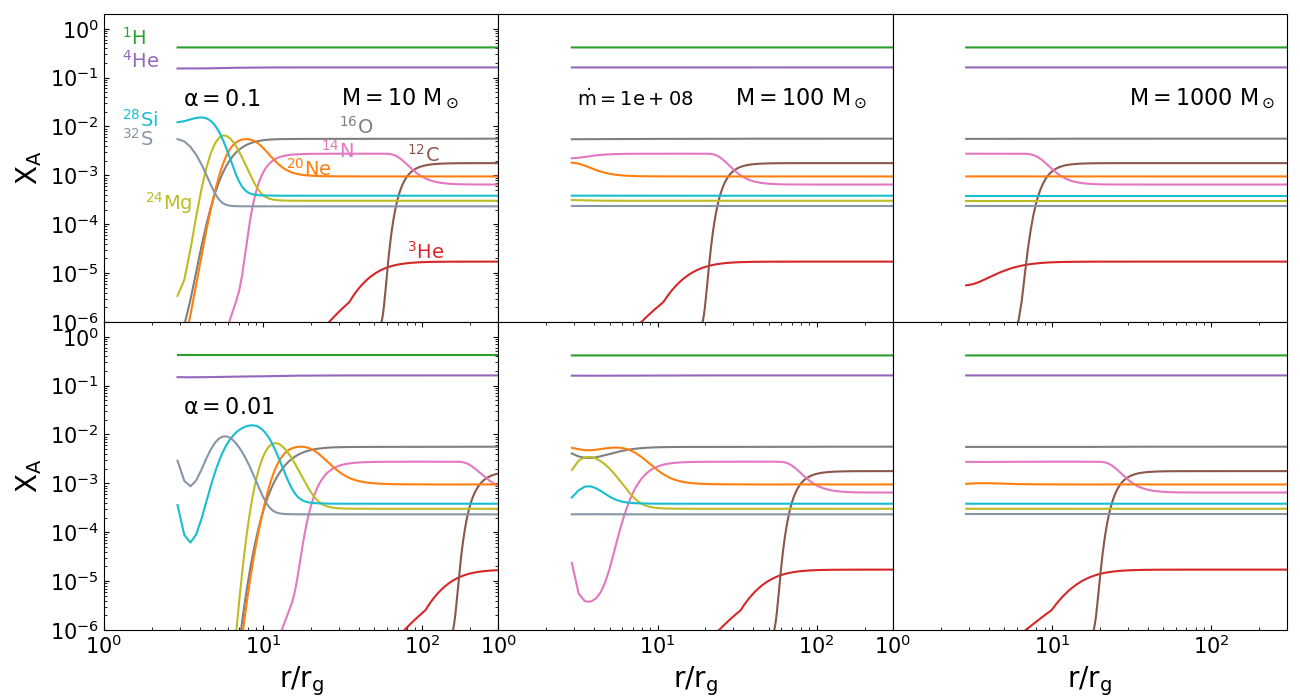}
    \caption{The radial distribution of the mass fraction $X_{\rm A}(r)$ for key isotopes are shown, for different black hole mass (from left to right panels), and differnt viscosity $\alpha$ (from top to bottom panels). The profiles are displayed at $\dot{m} = 10^8$. In the outer regions of the disk, nuclear burning is negligible, and the mass fractions closely resemble solar abundance. }
    \label{fig:mass_fraction}
\end{figure*}

\begin{figure}
    \center 
    \includegraphics[width=0.4\textwidth]{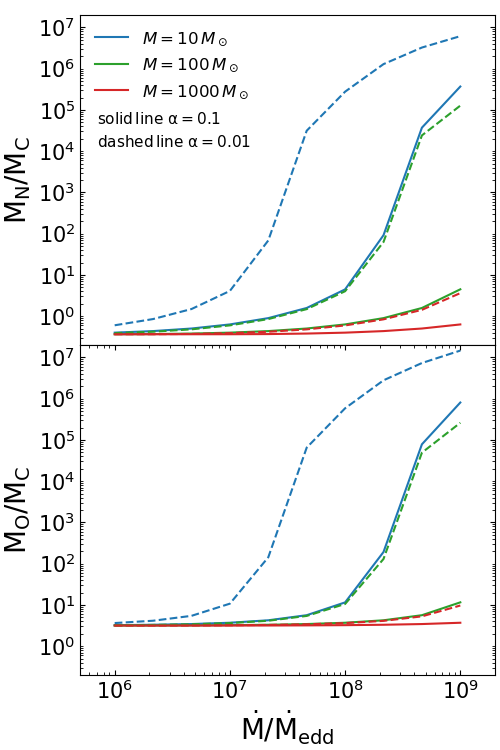}
    \caption{The relative ratios for the cumulative metal mass distribution within $100 r_{\rm g}$. The upper panel illustrates the ratio of enclosed nitrogen mass to carbon mass as a function of accretion rates, while the lower panels depict the ratio of oxygen to carbon.}
    \label{fig:metals}
\end{figure}

\section{Results}

\label{sec:results}

In our work, we choose a Kerr black hole and use the gravitational radius $r_{\rm g}$, to be the unit of the length scale. The BH spin $a$ is set as $a = 0.9$. We calculate the Eddington rate $\dot{M}_{\rm edd}$ as $\dot{M}_{\rm edd} = L_{\rm edd}/c^2$, where $L_{\rm edd}$ is the Eddington luminosity. In this section, we discuss the numerical results that we have obtained by solving the full equations.

\subsection{The  steady-state disc structure}
\label{sec:stabil}

We begin by examining the disc structure in a stationary accretion disc model. In Figure \ref{fig:disk_profile}, we present the profiles of density $\rho$, temperature $T$, radial/sound velocity $v_{\rm r}/c_{\rm s}$, and disk height aspect ratio $H/r$ at varying accretion rates,  for models with a black hole mass of $M = 100\, M_\odot$. The upper panels illustrate the profiles for $\alpha = 0.1$, while the lower panel corresponds to $\alpha = 0.01$. Generally, both temperature and density profiles increase towards the center. For models with lower accretion rates, the density distribution with respect to the radius closely resembles the profile $\rho \sim r^{-1.5}$, in line with the theoretical slim-disk models \citep{watarai06}. However, numerical models suggest a slightly flatter profile of $\rho \sim r^{-0.73}$ \citep{kitaki18}. It should be noted that we have not considered the impact of outflow above the disk in our analysis, which may contribute to the differences observed in the density profile. The temperature profile follows a trend $T \sim r^{-0.5}$, which is broadly consistent with the early theoretical expectations of $T \sim r^{-0.63}$ \citep{watarai06} and the numerical predictions of $T \sim r^{-0.54}$ \citep{kitaki18}.

When the Eddington rate ratio $\dot{m} \equiv \dot{M}/\dot{M}_{\rm edd}$ reaches $\dot{m} = 10^8$, our solutions reach a temperature as high as $T \sim 10^{9}\,\rm K$ in the innermost radii of the disc. Even for the rate ratio of $\dot{m}  = 10^6$, the temperature in the inner region has already exceeded the temperature thresholds for the CNO nuclear burning. 

In models with accretion ratios $\dot{m} \leq 10^8$, the density for $\alpha = 0.01$ (as shown in the lower panels) is $4$ to $6$ times greater than that in models with $\alpha = 0.1$, for the same accretion rate. While, the temperature differs by no more than approximately a factor of $2$. As the accretion ratio reaches $\dot{m} \geq 10^8$, nuclear burning becomes active, resulting in a shallower density and temperature profile within $\sim 100\, r_{\rm g}$. In systems with high accretion rates, advection cooling predominates over radiative cooling to counterbalance viscous heating, allowing the inner temperature of the disk to vary with different values of $\alpha$.

In the third panel, we present the radial velocity and sound speed relative to the speed of light as functions of radius. The gas flow passes through the sonic point around $2-3$ times the gravitational radius, which is close to $r_{\rm ms}$. The sound speed in the inner region can reach speeds as high as $0.2$ times the speed of light, and this remains relatively constant regardless of the accretion rate or the $\alpha$ values. When $\alpha = 0.01$, the angular momentum transport is ineffective, leading to an accumulation of flow in the outer region. This is evident as the radial velocity decreases by an order of magnitude compared to models where $\alpha = 0.1$.

The disk scale height is relatively large, approximately with $H/r \sim 1$. It is important to consider the vertical structure for a thick disk, which we have currently overlooked. It has been observed that the gas may inevitably move away from the surface of the disk when the radiation from the disk is strong, with a relative disk thickness aspect ratio $H/r \ge \sqrt{2}/2$ \citep{hubeny90,gu07,gu12,cao15,feng19,cao22}. Does the vertically modified structure affect the disk with nuclear burning? The disk temperature at the mid-plane is primarily determined by the energy balance, while the outflow, which removes the disk angular momentum and energy at the surface region, may have little impact on the temperature profile. However, as the angular momentum is carried away, the density along the vertical structure may be adjusted, potentially leading to an increase in density at the mid-plane \citep{gong17}. Consequently, nuclear burning could be more significant than initially anticipated.

In models where the black hole masses are $M = 10\, M_\odot$ and $M = 1000\, M_\odot$, we observe similar density profiles but with varying magnitudes. For example, the density for $M = 10\, M_\odot$ is approximately an order of magnitude greater. Meanwhile, the temperature profiles for various black hole masses, but at the same accretion rates, differ by merely a factor of two or three. The radial velocity, sound speed, and the ratio H/r exhibit minimal variation with respect to the black hole mass.

In Figure \ref{fig:ratio_qnuc}, we present the contribution of nuclear burning in the overall heating rate at a distance of $3\, r_{\rm g}$. The various colors indicate the proportional impact depending on the central black hole mass. The upper panel displays the profiles for $\alpha = 0.1$, while the lower panel illustrates the case for $\alpha = 0.01$. As the accretion rate rises, the significance of nuclear burning in the heating process also increases.

When $\alpha$ is smaller, the viscosity could not efficiently remove the gas angular momentum, leading to gas accumulation, reduced radial velocity, and higher density and temperature. Consequently, nuclear heating becomes more pronounced for lower values of $\alpha$. For example, when $\alpha = 0.01$ and $M = 10\, M_\odot$, the critical rate where nuclear burning dominates the total heating rate is approximately $\dot{m} = 10^7$. Conversely, for $\alpha = 0.1$, this critical rate is shifted to $\dot{m} = 10^8$. As the central black hole mass grows, the significance of nuclear burning may diminish due to the inverse relationship between the disk effective temperature and the black hole mass \citep{abramowicz88,watarai06,kato08}. Thus, we anticipate a more noticeable impact of nuclear burning in less massive black holes.

Various kinds of time variability in the light curves have been observed in black hole accretion systems, and these variations may stem from the instability of accretion disks \citep{hameury20}. An effective approach to investigating these instability processes is to analyze the $\dot{M} - \Sigma$ plane and the $T - \Sigma$ plane \citep{abramowicz88,kato08,yuan14}. These methods are highly effective for studying both disk secular and thermal instabilities, respectively.

We present the $\dot{M} - \Sigma$ plane and the $T - \Sigma$ plane at $r = 3 r_{\rm g}$ for various black hole masses in Figure \ref{fig:instablility}. Models with $\alpha = 0.1$ and $\alpha = 0.01$ are depicted as solid and dashed lines, respectively. In the event of instability, a turning point occurs in the curve, marking the critical stability point. This characteristic curve, known as the $S$-curve, has been widely used in the analysis of accretion-disk instability. However, as observed in Figure \ref{fig:instablility}, the accretion rate and temperature consistently show a positive correlation with the surface density, ensuring the stability of the accretion flow even when accounting for the contributions of nuclear burning.

Early studies by \citet{taam85} conducted a local analysis revealing the possibility of a phase transition in the disk, from a cold, low-viscosity state to a hot, high-viscosity state, due to the thermonuclear flash instability. It is important to note that the total heating energy is balanced by radiative cooling in this study, omitting advection cooling. In contrast, our research considers advection cooling, which efficiently dissipates excess heat and contributes to the stability of the disk.

Observations of novae and Type I X-ray bursts are believed to result from unstable nuclear burning on the surfaces of white dwarfs and neutron stars. However, in the accretion disk of a black hole, as depicted above, nuclear stability is observed. A significant distinction lies in the presence of a surface; the black hole lacks a surface, preventing mass accumulation in the inner region of the disk.


\subsection{Nucleosynthesis and disc composition}

The nuclear fusion process occurs in the inner region of the disk, leading to the burning of progressively heavier elements with decreasing radii. When calculating nucleosynthesis, the code retrieves the disk density and temperature data from the steady-state outcome. The calculation concludes at $1$ second, approximately matching the viscous time scale of the disk.

The radial distributions of the mass fraction $X_{\rm A}(r)$ for key isotopes are illustrated in Figure \ref{fig:mass_fraction}. The profiles are displayed at $\dot{m} = 10^8$, showcasing various black hole masses (from left to right panels) and different values of $\alpha$ (from top to bottom panels). In the outer regions of the disk, nuclear burning is negligible, and the mass fractions closely resemble solar abundance. 

In a model with $\alpha = 0.1$ and $M = 10 M_\odot$ (top left panel), at radii $r = 200 r_{\rm g}$, the temperature at the disk mid-plane reaches approximately $\sim 10^8\,\rm K$, triggering the conversion of initial carbon $\rm ^{12}C$ into $\rm ^{14}N$. Within $200\, r_{\rm g}$, the $\rm ^{12}C$ is significantly consumed. As the radius decreases to around $r \sim 10\, r_{\rm g}$, the temperature increases further, leading to the fusion of $\rm ^{14}N$ into $\rm ^{20}Ne, ^{24}Mg$, and $\rm ^{28}Si$. Closer to the center at $r < 5\, r_{\rm g}$, $\rm ^{20}Ne$ and $\rm ^{24}Mg$ undergo further fusion to form $\rm ^{28}Si$ and $\rm ^{32}S$. For larger black hole masses, yet with $\alpha = 0.1$, the mass fractions of $\rm ^{20}Ne, ^{24}Mg, ^{28}Si$ and $\rm ^{32}S$ exhibit minimal variation. 

The proportions of $\rm ^{1}H$ and $\rm ^{4}He$ remain largely unchanged, while $\rm ^{12}C$ and $\rm ^{3}He$ are significantly depleted in all models.

Early simulations of the super-critical accretion disk have shown that magnetic buoyancy and convection can transport both energy and mass to the disk surface, allowing a portion of the gas to be carried away by the outflow \citep{jiang14}. In our analysis, we can estimate the vertical convection timescale as $t_{\rm conv} \sim H/c_{\rm s}$, whereas the viscous timescale in the radial direction is $t_{\rm vis} \sim r/v_{\rm r}$. This results in $t_{\rm conv}/t_{\rm vis} \sim (H/r)(v_{\rm r}/c_{\rm s})$. As shown in Figure \ref{fig:disk_profile}, the cross-sonic point is approximately $3 r_{\rm g}$, and the ratio $H/r$ is around $1$. Therefore, outside the sonic point, the ratio $t_{\rm conv}/t_{\rm vis}$ might be less than $1$. Consequently, metals formed by nuclear burning could be carried into the outflow before reaching the horizon, thereby enriching the surrounding environment.

In the vicinity of the BLR, the gas has been observed to carry significant amounts of metals, often several times higher than the solar abundance \citep{hamann99,nagao06,shin17,lai22,huang23}. Assessing the abundance levels in the BLR presents difficulties, with collisionally excited line intensities in comparison to $\rm Ly\alpha$ showing limited sensitivity to metal content. Common indicators of chemical abundances involve ratios such as $\rm N\, V/C\, IV$ and $\rm  (Si\, IV + O\, IV)/C\, IV$ \citep{hamann99,huang23}.

Since most abundance indicators are based on carbon $\rm C$, which is a key component in our models. Burning of this carbon could potentially affect the line ratios observed. To simplify the analysis and avoid complex line ratio calculations due to photoionization and collisonal effects, we focus on presenting the mass content ratios, which still offer an approximate representation of the relative contributions. In Figure \ref{fig:metals}, we present the relative ratios for the cumulative distribution of the metal mass within $100\, r_{\rm g}$. The upper panel illustrates the ratio of the enclosed nitrogen mass to the carbon mass as a function of the accretion rates, while the lower panel depicts the ratio of oxygen to carbon. The comprehensive 3D numerical simulations conducted by \citet{jiang19} demonstrate that outflows originate from approximately $50$ gravitational radii when the accretion rate achieves $\dot{m} = 1500$. Given that our accretion rates exceed this, we compute the cumulative distribution of metal mass within $100\, r_{\rm g}$ for illustrative purposes.

It is crucial to acknowledge the possibility that the outflow rate might vary with the disk radius. Based on simulations performed by \citet{kitaki18}, the outflow rate stays roughly constant with respect to radius within the inner region ($r < 120 r_{\rm g}$). In contrast, the models presented by \citet{jiang19} show an increase in the outflow rates as the radius increases (refer to their Figure 7, indicated by the dashed black line). Thus, our estimations regarding the cumulative metal mass distribution provide merely an approximate assessment of metal enrichment.

In Figure \ref{fig:metals}, it is evident that the relative mass ratios increase as the accretion rates increase. In the case of models where $\alpha = 0.1$ (represented by solid lines) and the black hole mass exceeds $M = 100 M_\odot$, the ratio of $\rm M_N/M_C$ shows minimal variation for accretion rates below $\dot{m} = 10^7$, but undergoes a modest change by a few factors once the rate reaches $\dot{m} = 10^9$. Conversely, for black hole mass $M = 10 M_\odot$, the $\rm M_N/M_C$ ratio exhibits a variation spanning two to three orders of magnitude when $\dot{m} = 10^9$. A similar trend is observed for the $\rm M_O/M_C$ ratio. In particular, models with $\alpha = 0.01$ (depicted by dashed lines) yield considerably higher mass ratios in all scenarios. At an accretion rate of $\dot{m} = 10^9$, the ratios of $\rm M_N/M_C$ and $\rm M_O/M_C$ could be increased by up to $5 - 7$ orders of magnitude. It is worth mentioning that for all models with accretion rates below $\dot{m} = 10^7$, the carbon burning in the AMS disk has a negligible impact on the line ratios. 

Previous research conducted by \citet{wang23b} focused on studying how AMSs evolve in terms of population to generate metals, where the AMS is postulated to include a stellar component. This process involves metals being created through supernova explosions. In our study, we suggest that the AMS comprises a BH, and through nuclear burning within the AMS disk, the ratios of $\rm M_N/M_C$ and $\rm M_O/M_C$ could increase depending on the mass and accretion rate of the BH.  As carbon-depleted gas is expelled through outflows, variations in the mass ratio may manifest in the observed line ratios and resultant observed line-indicated metallicity. 


\section{Discussion and Concluding Remarks}

We have considered a stellar mass black hole embedded within an AGN disk, and studied the nuclear burning in the fast accretion disk onto this black hole. The structures of the accretion disks with Eddington ratio around $\dot{m} \sim 10^6 - 10^9$ have been solved. A number of conclusions can be drawn.

\begin{itemize}

\item For an accretion rate $\dot{m} \ge 10^6$, the mid-plane temperature in the inner region of the disk has already surpassed the thresholds required for nuclear burning. As the accretion rate increases, the importance of nuclear burning in the heating process also grows. This significance becomes more pronounced for lower viscosity values denoted by $\alpha$ and for less massive black hole mass.

\item The disk remains stable against the thermal and secular instabilities due to advection cooling counteracting the effects of nuclear heating. The lack of a solid surface around a black hole prevents an excessive accumulation of mass in the inner disk region.

\item Findings from nucleosynthesis point to significant burning of $\rm ^{12}C$ and $\rm ^{3}He$, especially in the context of black holes with masses around $10 M_\odot$ and accretion rates exceeding about $\sim 10^7$ times the Eddington rate. The ejection of carbon-depleted gas via outflows may lead to a rise in the mass ratio of oxygen or nitrogen to carbon. This change could be reflected in the observed line ratios like $\rm N\, V/C\, IV$ and $\rm O\, IV/C\, IV$. Consequently, these elevated spectral line ratios could be interpreted as indications of super-solar metallicity in the broad line region.

\end{itemize}

The ratio of ultraviolet $\rm Fe\, II$/$\rm Mg\, II$ in the BLR exhibits minimal cosmic evolution, possibly due to star formation occurring within the AGN nuclear region \citep{shin17,sarkar21}. The temperature threshold for Fe production is approximately $\sim 3 \times 10^9\,\rm K$, which remains too high for our current models \citep{prialnik09}. Although our models indicate partial depletion of Mg with $M = 10\,M_\odot$ (left panel of Figure \ref{fig:mass_fraction}), this could lead to an increase in the Fe/Mg ratio. However, because of the limited range of parameters, it is unlikely that AMS disk nuclear burning alone can explain all observed Fe/Mg lines.

The star formation models to explain the super-solar metallicity require a quite high star formation rate \citep{wang23b}. Estimating star formation rates directly from observations in such dense regions remains a challenge. Numerical simulations suggest that the actual star formation rate may not be as high as anticipated \citep{hopkins24}. This raises questions about the sole contribution of star formation to metal enrichment in the BLR. Nevertheless, if we consider the enrichment of depleted C or Mg gas from AMS nuclear burning disk, the required star formation rate could be lowered. Surely, a thorough investigation into the stellar population within the AGN disk is crucial to comprehend the impact of AMS on BLR metallicity \citep{wang23b,chen24}.

\section*{Acknowledgements}

The authors thank Xian Chen for helpful discussions on the general topics of this work and Wei-Min Gu for helpful comments on the transonic nature of the accretion flow. 
Y.L. acknowledges the support from NSFC grant No. 12273031. 
J.M.W. acknowledges financial support from the National Key R\&D Program of China (2021YFA1600404 and 2023YFA1607904), the National Natural Science Foundation of China (NSFC; 11833008, 11991050, 11991054, and 12333003).


\section*{Data Availability}

The data underlying this article will be shared on reasonable request to the corresponding author.
 


\bibliographystyle{mnras}
\bibliography{ms} 







\label{lastpage}
\end{document}